\documentclass[useAMS,usenatbib]{mn2e}

\usepackage{graphicx}
\usepackage{graphpap}
\usepackage{epsf}
\usepackage[utf8]{inputenc}
\usepackage{hyperref}
\usepackage[english]{babel}
\newcommand{\eprint}[1]{\href{http://arxiv.org/abs/#1}{#1}}
\newcommand{\adsurl}[1]{\href{#1}{ADS}}
\providecommand{\url}[1]{\href{#1}{#1}}

\newcommand{\araa}{{ ARA\&A~\/}}

\newcommand{\nat}{{ Nature~\/}}

\newbox\grsign \setbox\grsign=\hbox{$>$} \newdimen\grdimen
\grdimen=\ht\grsign
\newbox\simlessbox \newbox\simgreatbox \newbox\simpropbox
\setbox\simgreatbox=\hbox{\raise.5ex\hbox{$>$}\llap
     {\lower.5ex\hbox{$\sim$}}}\ht1=\grdimen\dp1=0pt
\setbox\simlessbox=\hbox{\raise.5ex\hbox{$<$}\llap
     {\lower.5ex\hbox{$\sim$}}}\ht2=\grdimen\dp2=0pt
\setbox\simpropbox=\hbox{\raise.5ex\hbox{$\propto$}\llap
     {\lower.5ex\hbox{$\sim$}}}\ht2=\grdimen\dp2=0pt

\title[Direct Distance Measurements to SN~2009ip]{Direct Distance Measurements to SN~2009ip}

\author[M. Potashov, S. Blinnikov, P. Baklanov, and A. Dolgov]
{
  M. Potashov$^{1}$\thanks{marat.potashov@gmail.com},
  S. Blinnikov$^{1,2,3}$\thanks{sergei.blinnikov@itep.ru},
  P. Baklanov$^{1,2}$\thanks{baklanovp@gmail.com},
  and A. Dolgov$^{1,2,4}$\thanks{dolgov@fe.infn.it}\\
  $^{1}$Novosibirsk State University, Novosibirsk 630090, Russia\\
  $^{2}$Institute for Theoretical and Experimental Physics, Moscow 117218, Russia\\
  $^{3}$Sternberg Astronomical Institute, Moscow State University, Moscow 119992, Russia\\
  $^{4}$University of Ferrara and INFN, Ferrara 44100, Italy\\
}

\begin{document}

\date{Accepted 2013 January 30. Received 2013 January 25; in original form 2012 December 31}

\pagerange{\pageref{firstpage}--\pageref{lastpage}} \pubyear{2012}

\maketitle

\label{firstpage}


\begin{abstract}
  We demonstrate the applicability of our new method (the Dense Shell Method or DSM)
  for the determination of astronomical distances by calculating the distance to SN~2009ip.
  The distance to this supernova has been accurately determined in the standard
  approach via the cosmic distance ladder and has been found to be 20.4~Mpc.
  Our direct method, assuming the most reasonable parameter values, gives a very close result,
  namely $\approx20.1\pm0.8$ ($68\%$~CL)~Mpc to SN~2009ip.
\end{abstract}



\section{Introduction}

In ref. \citep{Blinnikov2012} we introduced a new method for measuring cosmological distances,
which we suggest calling the Dense Shell Method (DSM).
The method relies on observations of an expanding dense shell in SN~IIn
and allows one to determine the linear size of such a shell in absolute units,
and hence the distance to it, without addressing the cosmological distance ladder.
Using this method \citet{Blinnikov2012} have calculated the distance to SN~2006gy.
But the result suffers from quite a large error
due to uncertainty in the interstellar extinction and in the color temperature.
Moreover, a small number of observations of the rising part of the light curve,
when the shell is believed not to fragment, also lead to a larger uncertainty in the measurements.
In this paper, we examine the applicability of our method to the case of the SN~2009ip,
which exploded again in 2012.
We show that the DSM allows one to obtain much more accurate results in this case.


\section{Direct distance determination by the DSM}
\label{sec1}

Supernovae of type IIn are observed at very large redshifts \citep{Cooke2008,Cooke2009,Moriya2012}.
Since they do not enter the coasting-free expansion phase,
neither expanding photosphere method \citep{KK1974}
nor spectra-fitting expanding atmosphere method \citep[SEAM,][]{BaronSEAM}
techniques are directly applicable.
Luminous Blue Variables can eject surface layers several times,
filling circumstellar space with expanding clouds of relatively dense matter.
The kinetic energy of those ejecta may be substantially lower than that of a genuine supernova
\citep[see][]{HegWoo2002}.
Narrow lines in the spectra of SN~IIn indicate that the matter of the first ejections had velocities
an order of magnitude lower than those of the ordinary supernovae
\citep[see][]{GraNad1986}.

SN~2009ip has produced several outbursts which have been observed directly.
The physical mechanisms of those ejections can be different \citep{HegWoo2002,Chevalier2012,Soker2012}.
The time spans between the outbursts can vary substantially.
If the time span is large enough, a cavity can be
formed between the star and the locus of the maximum density of the cloud.
If an LBV giant surrounded by such a cloud from the primary explosion
were to outburst again as a genuine supernova or to produce some other kind of a high-velocity outburst,
the new ejected material would have a much higher velocity.
So in the first days after the shock breakout,
the ejecta would propagate through the surrounding low-density cavity
of the cold and transparent cloud.
The spectra would contain only broad lines born in the high-velocity ejecta.
The ejecta would propagate and collide with high-density cloud layers.
As a result of this collision, shock waves would run in both directions from the locus of the collision:
forward (into the cloud) and backward (into the ejecta).

Below we will consider only spherically symmetric outbursts.
Our calculations show that the velocities of these shocks relative to the ejecta are low.
This is a consequence of the mass continuity
\citep[see equation~9 in][and discussion]{Blinnikov2012}.
One can consider the two shocks as ``glued together'' into a dense shell.
Moreover, all the kinetic energy of the matter accreted on to the dense shell is transformed into
electromagnetic radiation and escapes the cloud.
This is why we call this structure a cold dense shell (hereafter CDS).
Nevertheless, the part of the cloud in front of the CDS would be heated up
and would shine, producing radiation with narrow spectral lines.
For the small optical depth of the heated cloud, CDS may becomes visible
\citep{ChugaiEa04,WooBliHeg2007}.
In this case the observed spectrum would have narrow lines on a broad pedestal.
In addition, the matter of the cloud would be accumulated in the CDS, gradually breaking it,
so that eventually our assumption to relative shock velocities would not be justified anymore.
Future studies would need to investigate the stability of the CDS \citep[see][]{vanMarley2010}.

Calculations show that under certain conditions the photosphere is close to the CDS and
moves together with it
\citep[e.g., 2006gy,][]{WooBliHeg2007,Blinnikov2012}.
Assuming the equality of the velocities of the matter and the photosphere,
one can find the matter velocity from the broad spectral features
that allow measuring the photosphere velocity, $v_{\rm ph}$.
If a CDS has already been formed, the radius of the photosphere changes during the time interval $dt$,
evidently, as $d R_{\rm ph} = v_{\rm ph} dt$.
A set of models has been built that satisfactorily describes
a number of parameters of the ensuing process,
including the time dependence of the flux, the color temperature,
the cloud and ejecta velocities, and the cloud density.
All the models contain a free parameter, which
is the initial radius of the CDS shell.
Thus, they do not correspond to a certain distance to the star.
Further on, our DSM algorithm selects those models that have
the best fit for a sought-for distance.
Below for illustration we will consider only primitive models of the blackbody CDS.
Let us assume that observations of the supernova spectrum
are frequent enough to determine the variation of
the photosphere radius according to the relation
$dR_{\rm ph}=v_{\rm ph} dt$
for several time moments $t_i$ with $dt$ being the time interval between the measurements.
Let $\Delta R_i \equiv \int_{t_0}^{t_i} v_{\rm ph} dt$ be the increase of the radius during a large time interval
from the initial moment up to the $i$-th time moment.
We denote the initial radius (unknown to us) as $R_0$, and
$R_i \equiv R_0+\Delta R_i$ for $i=1,2,3, \ldots$.

Then using eq. equation~4 from~\citep{Blinnikov2012}, we find
\begin{equation}
  \zeta_{\nu i} (R_0+\Delta R_{i}) \sqrt{\pi B_\nu(T_{c \nu i})} = 10^{0.2A_\nu} D \sqrt{F_{\nu i}} ,
  \label{sqrtDistAv}
\end{equation}
where $\zeta_{\nu i}$ is the dilution factor,
$T_{c \nu i}$ is the color temperature, obtained from the spectrum,
function $B$ is the blackbody intensity,
$F_{\nu i}$ is the observed flux
and $A_\nu$ is the extinction in stellar magnitude units.
All quantities are defined for the frequency $\nu$.
And finally, $D$ is an unknown distance to the star.

A proper model would allow calculating a set of $\zeta_{\nu i}$
and $ T_{c \nu i}$ for all observational points.
From the measured flux, $F_{\nu i}$, and increments of the radius, $\Delta R_{i}$, we can find,
using the least-squares method, the initial radius, $R_0$, and the combination
$a_\nu \equiv 10^{0.4A_\nu} D^2$.
Instead of frequency $\nu$ one may use index $s$ labelling one of the broad-band filters.
To find the distance, $D$, we need to know $A_s$, which can be found from astronomical observations.

As we have noticed in ~\citep{Blinnikov2012}, the values of $\zeta$ do not vary too much
from model to model and only weakly depend on the photospheric radius.
In what follows, we assume that the dilution is absent and thus
the correction factor is unity, $\zeta=1$,
which is close to the real values of $\zeta$ with an accuracy of about $\sim10\%$ as it was found in our
radiation hydro models \citep{WooBliHeg2007,Moriya2013} for the growing part of the light curve.
Of course, a construction of a realistic, more accurate theory requires building a hydrodynamical
model not only for the light curve but also for the spectral line profiles with an account of the dilution
and projection effects as it has been done for the recent Cepheid models
\citep{Gautschy1987,Sabbey1995,Storm2011b,RastorguevDambis2011,RastorguevDambis2012}.


\section{Distance to the SN~2009\lowercase{ip}}
\label{sec2}

We took the observational data from \citep{Prieto2012a, Mauerhan2012, Pastorello2012}
and the web page \citet{Prieto2012b}.

SN~2009ip began to brighten rapidly after September 24,
possibly due to the interaction of the ejecta with the circumstellar material
[see \citet{Prieto2012a} and discussion in \citet{Mauerhan2012}].
Unfortunately, starting from September 28 the broad lines have mostly disappeared,
possibly due to an increasing opacity of the heated circumstellar matter,
so it has become difficult to determine the velocity of the CDS
(hereafter ``shell'' $=$ CDS).

We have used a short two-day period after September 24,
for which the velocity of the shell can be determined.
During this period the luminosity is proportional to the square of time \citep{Prieto2012a},
which corresponds to the constant expansion velocity and the color temperature of the photosphere.

The photometric data for SN~2009ip in the $R$ band \citep{Bessell2005} from all three papers are in good agreement
[see footnote 32 in \citet{Pastorello2012}].
We took the first 36 points listed in the table from the web page~\footnote{\url{http://www.astro.princeton.edu/~jprieto/sn09ip/photR.dat}}
in \citet{Prieto2012b}.
These points correspond to the days that are of most interest to us.

To estimate the expansion velocity $v$ of the shell we have
used the data obtained from the $H_{\alpha}$ absorption component.
During the epoch of a sudden increase in the luminosity of the SN~2009ip (around September 23-24)
\citet{Pastorello2012} and \citet{Mauerhan2012} indicate values
$v\approx13000$~km/s and $\approx13800$~km/s, respectively.
A full analysis is certainly needed to address a question, that has not yet been considered:
Are the layers of hydrogen, excited to the third level (which produces $H_{\alpha}$), close to the photosphere?
We have taken the matter velocity to be equal to $v\approx13400$~km/s.

We have adopted the extinction of $A_R=0.051$ mag following \citet{Mauerhan2012}.
In view of the smallness of this value, $10^{0.2A_R}$ is close to unity.
The error in $A_R$ has a minor effect on the output.
We have taken $0.001$ mag for the error value
but have not evaluated the impact of its changes on the final result.

\citet{Prieto2012a} obtained the black-body temperature in two different ways.
The first black-body fit took into account the data from the $R$ and $I$ filters and arrived to the value $T\approx14500$~K.
The second fit used the longer wavelength coverage (from the near--$UV$ to $V$--band) and the temperature obtained was
$T\approx19200$~K.
\citet{Prieto2012a} treated the derived values as effective temperatures,
but the method of obtaining them corresponded to color temperatures.
In our method these values are used as color temperatures.

To estimate the confidence interval of the calculated distance we have applied a resampling Monte Carlo (MC)
simulation based on these data.
We have resampled the values of $T$, the stellar magnitude, $m_R$, in standard filter $R$,
the reddening, $A_R$, and velocity, $v$, each with normal distribution.
For some of those quantities we have taken the standard deviations from the available data,
or fixed them ``manually'' for illustration, if they are not known.
For example, since we do not know the error of $v$ and $T$,
we have examined how the variations in their errors affect the response.

For obtaining the confidence intervals for the mean and median distances
it was sufficient to make $10^5$ MC tests.

Taking the temperature of the first \citet{Prieto2012a} estimate as $T\approx14500$~K
and the relative errors in $v$ and $T$ equal to $5\%$
we have obtained both the
mean and median distances $D\approx16.1$~Mpc
with a $68\%$ confidence interval $\pm0.6$~Mpc.

If we artificially increase the relative errors of $v$, $T$ by a factor of 2, we find
the distance $D\approx15.1$~Mpc
with a $68\%$ confidence interval $(-1.1,+1.2)$~Mpc.
The result is more sensitive to the error in the temperature.

\begin{figure}
  \centering
  \includegraphics[width=0.45\textwidth]{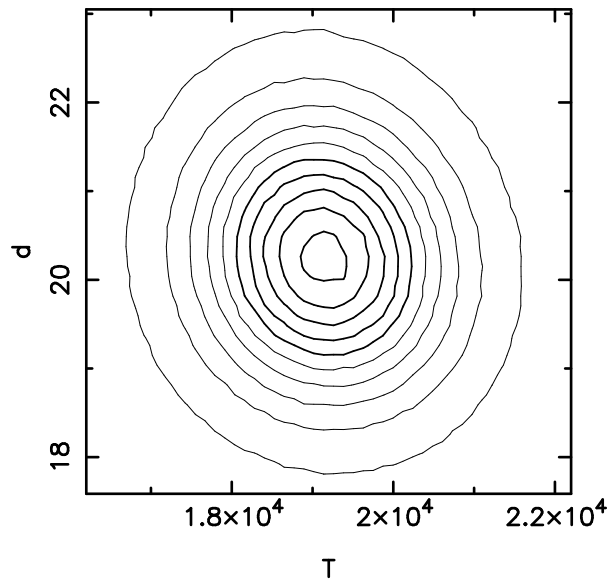}
  \caption
  {
    Monte Carlo resampling simulation of the distance $D$ to SN~2009ip by DSM
    with respect to temperature $T$.
    The isocontours of the probability distribution function (pdf) are shown with equal step in pdf.
    The observations from \citep{Prieto2012a, Mauerhan2012, Pastorello2012} have been used for 36 different time points
    from the \citet{Prieto2012b} web page.
    \label{pld09ip}
  }
\end{figure}

Taking the temperature of the second \citet{Prieto2012a} estimate as
$T\approx19200$~K and the relative error in $v$ and $T$ equal to $5\%$
we have obtained the distance $D\approx20.1$~Mpc
with a $68\%$ confidence interval $\pm0.8$~Mpc.
For an illustration of this case, we have built a slice of multi-dimensional probability density
functions on a ($T$-$D$)--plane.
The plot in Fig.~\ref{pld09ip} is built with $10^7$ samples to obtain better statistics near
the top of the distribution.
The largest thick-line contour in Fig.~\ref{pld09ip} is close to one standard deviation.

If we increase the relative errors of $v$ and $T$ by a factor of 2 again, we can find
the distance $D\approx19.9$~Mpc
with a $68\%$ confidence interval $(-1.4,+1.5)$~Mpc.
The answer is again more sensitive to the error in the temperature.
In all the cases the error rise leads to smaller median and mean distances.

These numerical experiments show that the results are rather robust given the level of accuracy of
the data and models.

The distance modulus for the galaxy NGC~725, where the SN~2009ip is situated,
is generally accepted as $\mu = 31.55$~mag \citep{Smith2010b}
which corresponds to 20.4~Mpc.
More detailed and complete studies (like SEAM) without the black-body approximation
but with the assistance of a complete hydrodynamical model
of the whole sequence of SN~2009ip outbursts
permit us to determine the temperature quite accurately.
However, we can already say now that in the frameworks of our models,
the temperature estimate $T\approx14500$~K is unsatisfactory
(if one believes in the generally accepted value of the distance).
The goal of the current paper is not to obtain the most reliable distance value
but to show that it can be easily obtained with reliable data.
Moreover, increasing the number of observations would reduce
the error in the final result as the square root of the number of observations.

Since the host galaxy NGC~725 is relatively close,
the Hubble parameter determined only from the redshift measurements
can be incorrect because of a large peculiar velocity of the galaxy.
On the other hand, the distance to the galaxy NGC~725 calculated with
the generally accepted value $H_0 = 71~\mbox{km/s/Mpc}$
can be significantly different from the distance obtained by other methods.
For instance, \citet{Levesque2012} uses the distance to NGC~725 equal to 24~Mpc.

The value of the distance is obtained here by a new direct method which
does not rely on the cosmic distance ladder!
To check the robustness of our results it is necessary to study
the effects of variations in correction factors in different SN~2009ip models.
We present here the values for the distance only as an illustration of the efficiency
of the proposed method.


\section{Conclusions}

We have obtained the distance to SN~2009ip equal $D\approx20.1\pm0.8$ ($68\%$~CL)~Mpc.
The relative error is much smaller than that in the case of SN~2006gy \citep{Blinnikov2012}.
If the observations of a supernova are of good quality,
and have a large number of data points,
the internal error of our method for such a supernova will be very small.
A comprehensive application of the method requires a complete supernova SEAM-like model
with complex hydrodynamics taking into account the entire
sequence of previous outbursts.
One should also evaluate the correction factors (e.g. dilution factor)
to eliminate possible systematic errors.

The distance determined in this paper is calculated under the assumption of
the spherically symmetric explosion.
The fact that it coincides with the generally accepted value shows
that the initial epoch of the growth of the SN~2009ip luminosity
is well described with an expanding spherical shell.

Constraining cosmological parameters and understanding the nature of
Dark Energy depend strongly on accurate measurements of distances in the Universe.
Our results on SN~2009ip support the conclusion that SNe~IIn may be used for cosmology
as \emph{primary distance indicators} with the new DSM method.


\section*{Acknowledgements}

SB and MP are grateful to Alexander Bondar for
discussions and encouragement
in 2012 September and October,
when the luminosity of the SN~2009ip
exhibited large variations.

The work is partly supported by the grants
of the Government of the Russian Federation (No 11.G34.31.0047),
by RFBR 10-02-00249, 10-02-01398,
by RF Sci.~Schools 5440.2012.2, 3205.2012.2,
and by a grant IZ73Z0-128180/1 of the Swiss National Science
Foundation (SCOPES).


\bsp

\label{lastpage}

\end{document}